# Generation of Flying Electromagnetic Donuts

N. Papasimakis[1,*], T. A. Raybould[1], V. A. Fedotov[1], D.P.Tsai[2], I. Youngs[3], and N. I. Zheludev[1,4]

[1]*Optoelectronics Research Centre & Centre for Photonic Metamaterials, University of Southampton, Highfield SO17 1BJ, UK*

[2]*Department of Physics, National Taiwan University and Research Center for Applied Sciences, Academia Sinica, Taipei 10617, Taiwan*

[3]*Physical Sciences Department, DSTL, Salisbury SP4 0JQ, UK*

[4]*Centre for Disruptive Photonic Technologies, School of Physical and Mathematical Sciences and the Photonics Institute, Nanyang Technological University, 637371, Singapore*

[*] n.papasimakis@soton.ac.uk

**Abstract:** Transverse electromagnetic plane waves are fundamental solutions of Maxwell's equations. It is less known that a radically different type of solutions has been described theoretically, but has never been realized experimentally, that exist only in the form of short burst of electromagnetic energy propagating in free-space at the speed of light. They are distinguished from transverse waves by a donut-like configuration of electric and magnetic fields with a strong field component along the propagation direction. Here, we report that such "Flying Donuts" can be generated from conventional pulses using a singular metamaterial converter designed to manipulate both the spatial and spectral structure of the input pulse. The ability to generate Flying Donuts is of fundamental interest, as they shall interact with matter in unique ways, including non-trivial field transformations upon reflection from interfaces and the excitation of toroidal response and anapole modes in matter, thus offering new opportunities for telecommunications, sensing, and spectroscopy.

## Introduction

Electromagnetic waves propagating in free-space are described by the source-free Maxwell's equations. Examples include the vast majority of waveforms employed in optical science, such as plane waves, Bessel beams, X-waves, Airy and Gaussian pulses. The electromagnetic fields of such solutions are typically transverse to the propagation direction, and they can be written as a product of space- and time-dependent functions. However, there is a very different family of exact solutions to Maxwell's equations, where the spatial and temporal structure are inherently linked and cannot be separated. Early work by Brittingham introduced the so-called focus wave modes, pulses that propagate without diffraction but carry infinite energy (1). Soon after Richard Ziolkowski addressed this issue by considering superpositions of the original focus wave modes resulting in finite energy pulses, termed Electromagnetic Directed Energy Pulse Trains (EDEPTs) (2, 3). This family of pulses and its derivatives include a wide-range of exotic waveforms, such as the modified power spectrum pulse (2), non-diffracting pulses with azimuthal dependence (4, 5), "focused pancake" pulses (6–8), and the "Flying Donut" (FD) pulses (9). The latter, in addition to their single-cycle nature, are distinguished by a donut-like configuration of electric and magnetic fields, which feature strong longitudinal field components parallel to the pulse propagation direction (see Fig. 1a). Importantly, owing to the unique spatiotemporal coupling, such pulses are exceptionally broadband. Potential applications of FD pulses include particle acceleration (9), while non-trivial interactions with interfaces have been predicted (10). In particular, owing to the topological similarity with the toroidal dipolar excitation and the presence of longitudinal field components, it has been shown that FD pulses can excite strong toroidal modes in dielectric particles, as well as anapoles, namely non-radiating combinations of electric and toroidal dipoles (10).

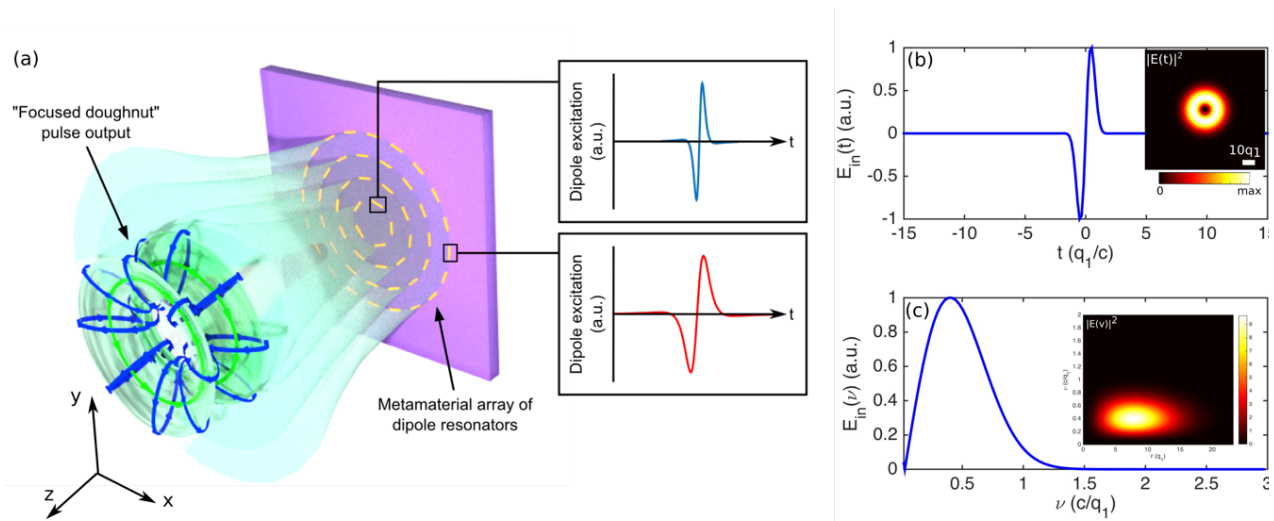

*Figure 1. (a) Schematic of a metamaterial generator of focused donut pulses. Concentric rings of azimuthally oriented linear dipole resonators with spatially and frequency dependent scattering properties lead to the emission of a toroidal single-cycle "flying donut" pulse. The resonators in different rings exhibit a different frequency dispersion (see insets). (b-c) The temporal (b) and frequency (c) form of the single-cycle radially polarized pulse driving the metamaterial generator. Inset to (b) shows the electric field intensity profile of the input pulse in the xy-plane (transverse to the propagation direction). Inset to (c) shows the spectral power of the input pulse at different distances from the pulse centre.*

However, due to the complexity of the electromagnetic field configuration, broad bandwidth, and complex spatial and temporal coupling, FD pulses have not been realised to date. Early works have suggested various generation schemes based on antenna arrays, where each element in the array is considered to be driven by an electrical signal with different amplitude and temporal shape (11, 12). In this case, the electromagnetic fields radiated by all antennas in the array combine in the far-field forming the desired pulse. Nevertheless, such configurations are complex and cannot be easily translated to optical frequencies. Here, drawing on recent advances in the field of metamaterials for dispersion (*13, 14*) and wavefront control (*15–17*), we introduce a method for the generation of FDs and other exotic pulses with complex spatiotemporal structure, in general.

## Results

The FD pulse exists in two different forms termed transverse electric (TE) and transverse magnetic (TM). In the case of a TE FD pulse propagating along the z-axis and coming into focus at $z=0$, $t=0$, the electric and magnetic fields are given by (in the cylindrical coordinates $\rho, \theta, z$) (*9*):

$$E_\vartheta = -4if_0\sqrt{\frac{\mu_0}{\epsilon_0}}\rho\frac{q_1 + q_2 - 2ict}{\left(\rho^2 + \left(q_1 + i(z-ct)\right)\left(q_2 - i(z+ct)\right)\right)^3} \quad (1)$$

$$H_\rho = 4if_0\rho\frac{-q_1 + q_2 - 2iz}{\left[\rho^2 + \left(q_1 + i(z-ct)\right)\left(q_2 - i(z+ct)\right)\right]^3} \quad (2)$$

$$H_z = -4f_0\frac{\rho^2 - \left(q_1 + i(z-ct)\right)\left(q_2 - i(z+ct)\right)}{\left[\rho^2 + \left(q_1 + i(z-ct)\right)\left(q_2 - i(z+ct)\right)\right]^3} \quad (3)$$

where $f_0$ is an arbitrary normalisation constant. The parameters $q_1$ and $q_2$ have the dimensions of length and represent respectively the effective wavelength of the pulse and the focal region depth. Beyond the focal region ($z>q_2$), the FD diffracts in the same manner as a Gaussian pulse with wavelength $q_1$ and Rayleigh length $q_2/2$. Owing to the duality of Maxwell's equations in free-space, the TM solutions are readily obtained by interchanging electric and magnetic field components. As it is evident from Eqs. (1-3), the FD pulse is a non-separable solution to Maxwell's source-free equations meaning that it cannot be written as a product of a space-dependent and a time-dependent function. Here, we focus on the generation of TE FD pulses.

Generating FD pulses presents a number of challenges. First, the broad frequency spectrum of the pulse requires broadband radiating elements. Second, the frequency content of the pulse varies substantially along the radial direction. We address these challenges by introducing a new class of gradient metamaterials that allow simultaneous spatial and temporal wavefront manipulations and are based on low Q-factor dipole resonators. We employ a cylindrically symmetric metamaterial array (see schematic representation in Fig. 1a), reflecting the symmetry of the FD pulse. The array consists of azimuthally oriented electric dipoles arranged in concentric rings. The spatiotemporal coupling is provided by varying the properties of the metamaterial elements across the radial direction according to the parameters of the targeted FD pulse (effective wavelength, $q_1$, and Rayleigh length, $q_2$). We consider illumination by an ultrashort Gaussian, azimuthally polarized pulse and calculate numerically the electromagnetic fields emitted by the array (see Methods). Such radially polarized pulses can be readily generated in the optical range in a number of ways, including beam interference (*18*), segmented waveplates (*19*) and, more recently, by dielectric metasurfaces (*20*).

Following the procedure outlined in the Methods section, we target an FD TE pulse with effective wavelength, q1, Rayleigh length, $q_2=300q_1$ and focused at $z_0=0$ and $t_0=0$. The results of our calculations are presented in Fig. 2, where the azimuthally polarized electric field is presented both for the ideal (target) FD pulse and the metamaterial-generated pulse at three different moments in time. As shown in Fig. 2a, shortly after the excitation of the metamaterial surface with the radially polarized pulse, a few-cycle pulse is emitted with a wavefront that feature strong side lobes and a "tail" close to the center of the pulse. However, upon propagation these inhomogeneities of the wavefront are rapidly damped due to diffraction (Fig. 2b), and after a propagation distance of $z=200q_1$, the pulse acquires a toroidal shape consisting of a single cyle (Fig. 2c). Moreover, the values of the electric field intensity for propagation distances between $z=100q_1$ and $z=200q_1$ remain almost constant, indicating the weakly diffracting behaviour of the generated pulse. As the propagation distance increases, the electric field profile of the generated pulse (Figs. 2a-c) converges to that of the target pulse (Figs. 2d-f) with both pulses diffracting in a similar way. An important difference here, however, is that the ideal FD pulse experiences shape changes due to Gouy phase shifts, with the pulse duration increasing from 1 to 1 ½ cyclea. Such changes are much less pronounced in the case of the generated pulse.

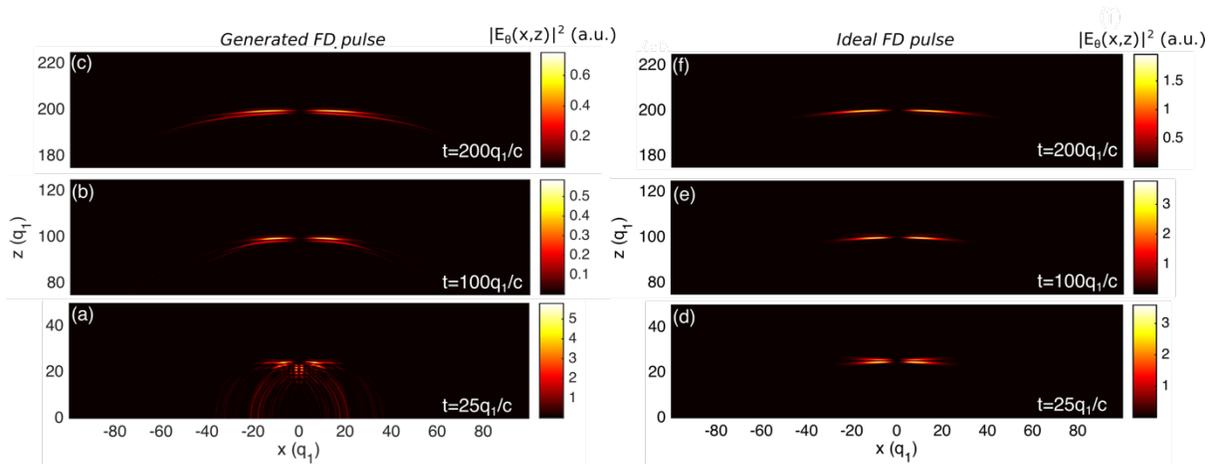

*Figure 2. Snapshots of the generated (a-c) and ideal (d-f) FD pulse at three different moments in time ($t=25q_1/c$, $100q_1/c$ & $20q_1/c$) with $q_2=300q_1$. Colormaps show the modulus squared of the azimuthal (normal to the xz-plane) electric field.*

One of the fascinating properties of the FD pulse is its space-time non-separability, which leads to a spatially varying frequency spectrum. This is in sharp contrast to the case of the input driving pulse, which is space-time separable and its frequency spectrum is uniform throughout the pulse wavefront. In our approach, the space-time non-separability is induced by the gradient structure of the metamaterial generator. The frequency spectra of the generated and target pulses are shown in Fig. 3 at three different propagation distances. Close to the metamaterial generator (Fig. 3a), the frequency spectrum of the generated pulse deviates significantly from that of the target pulse (Fig. 3d) due to presence of strong near-field contributions. However, the space-time non-separability can already be seen in the central part of its spectrum, which begins to resemble the spectrum of the target FD. As the pulses propagate further ($z=100q1$), the similarity between their spectra increases (Figs. 3b & 3e), although, due to diffraction, the side lobes of the generated pulse can still be clearly observed at the periphery. At a propagation distance of $z=200q_1$, these sidelobes are all but absent and the generated pulse assumes a form very close to that of an FD pulse. Here, the peak frequency, $v_{max}$, in the spectrum of the generated pulse is strongly position dependent varying from $v_{max} \cong 0.7c/q_1$ close to the centre of the pulse to $v_{max} \cong 0.2c/q_1$ at the outer areas of the pulse.

We would like to note that the metamaterial FD generator does not introduce any changes in the topology of the pulse: both the input radially polarized pulse and the output FD pulse exhibit a singularity at their respective centers. Rather, the main role of the FD generator is to introduce coupling of the spatial and temporal structure of the pulse.

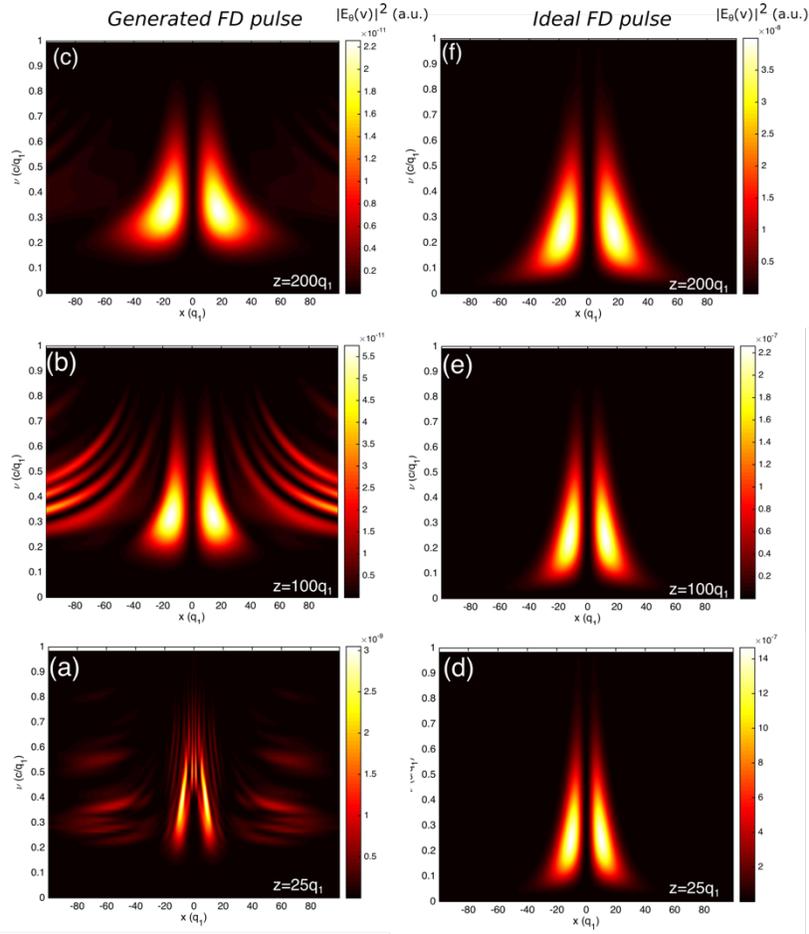

*Figure 3. Spatially varying spectra of the generated (a-c) and ideal (d-f) FD pulse at three different distances from the metamaterial generator ($z=25q_1$, $100q_1$ & $200q_1$) with $q_2=300q_1$. Colormaps show the modulus squared of the azimuthal (normal to the xz-plane) electric field.*

To quantify the similarity of the generated FD pulse to is target form, we introduce a frequency dependent figure of merit (FOM) based on the spatial overlap of the azimuthal electric fields of the two pulses:

$$FOM(\nu) = \left| \frac{\int E_{gen}(\nu) E_{ideal}(\nu) \, dr^3}{\int |E_{gen}(\nu)| dr^3 \int |E_{ideal}(\nu)| dr^3} \right| \quad (4)$$

Owing to the normalization of FOM, values close to null indicate low similarity between the pulses, whereas values close to unity indicate good match between the pulses. In Fig. 4a, we present FOM at three different propagation distances, corresponding to the frequency spectra in Figs. 3a-c. In the case of strong spectral mismatch, FOM is limited to relatively low values close the metamaterial generator, especially for high frequency components. This is a direct result of the appearance of side lobes, which significantly distort the wavefront of the generated pulse. However, at longer propagation distances FOM exceeds 0.7, indicating a high degree of topological similarity between the generated and target pulses. To further demonstrate that the generated pulse resembles closely the target FD pulse (rather than just an FD pulse), we performed a parametric scan, calculating the spatial overlap between the two pulses while varying the values of $q_1$ and $q_2$. As shown in Fig. 4b, the highest values of FOM occur close to the parameters of the target FD, hence validating our approach.

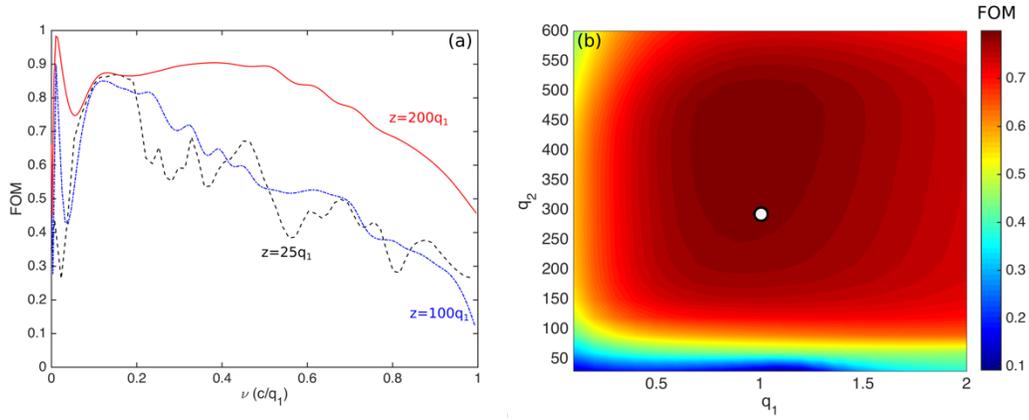

*Figure 4. (a) Frequency dependent figure of merit (FOM), defined as the spatial overlap of the generated pulse and the ideal (target) pulse at each frequency (Eq. (4)). Dashed black, dash-dot blue and solid red lines correspond to three different propagation distances $z=25q_1$, $100q_1$ & $200q_1$, respectively. (b) FOM calculated and averaged over all frequencies for different $q_1$ and $q_2$ parameters of the ideal FD pulse. The white circle marks the values of $q_1$ and $q_2$ parameters initially targeted in our study.*

## Discussion

In conclusion, we have introduced a new type of metamaterials that allows complex spatiotemporal manipulations of electromagnetic waves and have demonstrated numerically the generation of Flying Donut pulses. Our approach is based on metamaterial arrays of electric dipole resonators with orientation, resonance frequency and linewidth that vary spatially throughout the array, and hence FD generators can be readily realized with dielectric or plasmonic metasurfaces. FD pulses constitute a new type of broadband information carrier confined both in time and space, and as such they can be employed in telecommunications applications. Here, the non-separability of the FD pulse time- and space-dependence should allow the transmission of information over FD-specific channels, which can be distinguished from the conventional case of space- and time- separable pulses. Moreover, the space-time non-separability property of FD pulses enables the encoding of information about the spectral (or, equivalently, temporal) form of the pulse, into its wavefront, and vice-versa. This opens up intriguing properties for the spectroscopy and characterization of materials properties, where one could acquire spectral information simply through imaging the pulse wavefront, eliminating the need for spectrometers or monochromators. Conversely, information about the spatial shape of the pulse can be obtained through spectral measurements. Finally, such pulses can be considered as the propagating counterparts of the recently observed localized toroidal excitations in matter (*21*). They have also been shown to excite toroidal dipolar modes in dielectric particles (*10*), providing thus the basis for toroidal-specific forms of spectroscopy.

## Methods

### Input pulse
The metamaterial generator is driven by a single-cycle, azimuthally polarised pulse propagating in free-space, which provides the required field topology and bandwidth for the generation of FD pulses. Such free-space propagating pulses are space-time separable and can be described approximately by (*18*):

$$E_{in}(r,t) = \hat{r} E_0 R(r) T(t) \qquad (5)$$

with

$$R(r) = \frac{re^{-\left(\frac{r}{s}\right)^2}}{s} \tag{6}$$

$$T(t) = e^{i\omega_{in}t} e^{-\frac{\ln 2(2t/\tau_{in})^2}{2}}, \tag{7}$$

where $E_0$ is an arbitrary constant defining the energy of the pulse, $s$ is the waist, $\omega_{in}$ is the carrier frequency, and $\tau_{in}$ is the duration of the pulse. The waist of the pulse is defined such that it is equal to that of the target FD pulse. The carrier frequency is matched to the peak frequency of the target FD pulse, while the time duration of the input pulse is chosen to provide a single-cycle pulse with a bandwidth close to the FD pulse bandwidth. The transverse profile of the input pulse is shown in the inset to Fig. 1b. In a frequency representation, the input pulse can be written as:

$$E_{in}(r,\omega) = \hat{r} E_0 R(r) \Omega(\omega) \tag{8}$$

with

$$\Omega(\omega) = \int e^{i\omega t} T(t) dt \tag{9}$$

**Metamaterial pulse transformer**
The pulse generator for TE FD pulses consists of an array of 217 azimuthally polarized linear dipole scatterers, with prescribed frequency dependent radiation patterns, arranged in six concentric rings (see schematic in Fig. 1a). The metamaterial structure is cylindrically symmetric with dipole scatterers in the same ring being identical. In particular, dipoles in ring of radius $R_i$ are characterized by a frequency dependent complex polarizability $a_i(\omega)$. Here we are concerned with $\lambda/2$-resonators, hence their spectral response can be described by a Lorentzian centred at frequency $\omega_{o,i}$ with a decay rate $\gamma_i$. These parameters can be controlled by the length and width of the dipole resonators. The peak frequency and bandwidth of the dipole scatterers vary radially following the profile of the target FD pulse. As a result, the dipoles near the center emit shorter pulses at higher frequencies (Fig. 1b) compared to the dipoles at the periphery of the array (Fig. 1c). Our calculations are based on a point dipole model, where each resonator in the array is replaced by a point dipole scatterer with the same resonant polarizability.

A dipole at position $R_i$ in the array is excited by the electric field of the input radially polarised pulse at this site, $E_{in}(R_i,\omega)$. The associated electric dipole moment is:

$$p_i(\omega) = E_{in}(R_i,\omega) a_i(\omega) \tag{10}$$

At each frequency, $\omega$, the field scattered by dipole $p_i(\omega)$ at an observation point $R_o$ is calculated as:

$$E_i(\omega, R_o) = p_i(\omega) e^{-i\omega t} \frac{e^{-ik|R_i - R_o|}}{|R_i - R_o|} \tag{11}$$

The total field at point $R_o$ scattered by all dipoles in the array is:

$$E(\omega, R_o) = \sum_i E_i(\omega, R_o) \tag{12}$$

After a Fourier transform, we obtain the scattered pulse in the time domain in the form:

$$E(t, R_o) = \frac{1}{2\pi} \int e^{-i\omega t} E(\omega, R_o) d\omega \tag{13}$$

# Acknowledgements

The authors acknowledge the financial support of DSTL (DSTLX-1000068886), the Engineering and Physical Sciences Research Council (United Kingdom) (EP/G060363/1 and EP/M008797/1), the Leverhulme Trust, and the Ministry of Education (Singapore) (MOE2011-T3-1-005). The authors would like to thank V. Savinov and E. T. Rogers for numerous fruitful discussions. The authors declare that they have no competing interests.